%% file: main.tex
\documentclass[10pt,conference]{IEEEtran}
\IEEEoverridecommandlockouts

\usepackage{cite}
\usepackage{amsmath,amssymb,amsfonts}
\usepackage{algorithmic}
\usepackage{graphicx}
\usepackage{textcomp}
\usepackage{xcolor}
\usepackage{footnote}
\def\BibTeX{{\rm B\kern-.05em{\sc i\kern-.025em b}\kern-.08em
    T\kern-.1667em\lower.7ex\hbox{E}\kern-.125emX}}

\input{utils/macros}
\begin{document}
\bstctlcite{IEEEexample:BSTcontrol}

\title{ 
How Propense Are Large Language Models at Producing Code Smells? A Benchmarking Study 
}

\author{
\IEEEauthorblockN{
Alejandro Velasco$^1$, Daniel Rodriguez-Cardenas$^1$, Luftar Rahman Alif$^2$, David N. Palacio$^1$, Denys Poshyvanyk$^1$}
\IEEEauthorblockA{
$^1$Department of Computer Science, William \& Mary \\
$^2$Department of Software Engineering, University of Dhaka \\
\{svelascodimate, dhrodriguezcar\}@wm.edu, bsse1120@iit.du.ac.bd, \{danaderpalacio, dposhyvanyk@\}wm.edu}
}


\maketitle

\input{sections/0_abstract}

\input{sections/1_introduction}

\input{sections/2_benchmark}

\input{sections/3_case_study}

\input{sections/4_related}

\input{sections/5_future_plan}


\bibliographystyle{IEEEtran}
\bibliography{main}

\end{document}

%% file: utils/macros.tex
\usepackage{caption}
\usepackage{blindtext}
\usepackage{tcolorbox}
\usepackage[final]{pdfpages}
\usepackage{lipsum,multicol}
\usepackage{xcolor}
\usepackage{tikz}
\usepackage{listings}
\usepackage{enumitem}
\usepackage{hyperref}
\usepackage{amsfonts}
\usepackage{wrapfig}
\usepackage{subcaption} 
\usepackage{adjustbox}
\usepackage{colortbl}
\usepackage{fancybox}
\usepackage{multirow}
\usepackage[normalem]{ulem}
\useunder{\uline}{\ul}{}
\usepackage{enumitem}

\newcommand{\ie}{\textit{i.e.,}\xspace}
\newcommand{\eg}{\textit{e.g.,}\xspace}

\newcommand{\etal}{et al.\xspace}

\newtcolorbox{boxK}{
    fontupper = \small,
    sharpish corners, 
    boxrule = 0pt,
    toprule = 0pt, 
}

\newcommand{\equref}[1]{Eq.~\ref{#1}\xspace}
\newcommand{\secref}[1]{Sec.~\ref{#1}\xspace}
\newcommand{\figref}[1]{Fig.~\ref{#1}\xspace}
\newcommand{\tabref}[1]{Table~\ref{#1}\xspace}

\newcommand{\metric}{\textit{PSC}\xspace}
\newcommand{\dataset}{\textit{CodeSmellData}\xspace}
\newcommand{\benchmark}{\textit{CodeSmellEval}\xspace}

\newcommand{\llms}{\textit{LLMs}\xspace}
\newcommand{\llm}{\textit{LLM}\xspace}

%% file: sections/0_abstract.tex
\begin{abstract}
Large Language Models (\llms) have shown significant potential in automating software engineering tasks, particularly in code generation. However, current evaluation benchmarks, which primarily focus on accuracy, fall short in assessing the quality of the code generated by these models, specifically their tendency to produce code smells. To address this limitation, we introduce \benchmark, a benchmark designed to evaluate the propensity of \llms for generating code smells. Our benchmark includes a novel metric: \textit{Propensity Smelly Score (\metric)}, and a curated dataset of method-level code smells:  \dataset. To demonstrate the use of \benchmark, we conducted a case study with two state-of-the-art \llms, CodeLlama and Mistral. The results reveal that both models tend to generate code smells, such as \textit{simplifiable-condition} and \textit{consider-merging-isinstance}. These findings highlight the effectiveness of our benchmark in evaluating \llms, providing valuable insights into their reliability and their propensity to introduce code smells in code generation tasks.
\end{abstract}


%% file: sections/1_introduction.tex
\section{Introduction}
\label{sec:introduction}

{Large Language Models (\llms) have been used to automate multiple software engineering (SE) tasks, including code completion \cite{9616462, lin_when_2024, 7180092}, code summarization \cite{devanbu}, program repair \cite{tuffano_llms_repair}, clone detection \cite{10.1145/2970276.2970326}, and assertion generation \cite{Watson:ICSE20}. LLMs' applications have expanded from classification tasks (\eg defect detection, requirements triage) to generative tasks, \ie sequence synthesis in which \llms generate text or code from a given prompt \cite{DL-survey}. Ergo, ensuring LLMs' prediction quality is crucial, particularly for SE tasks that require rigorous evaluation due to the complexity of source code. In this work, we evaluate the propensity of \llms to generate or predict \textit{code smells}.}


{Code smells are symptoms of poor design and implementation choices, negatively affecting code comprehensibility and maintainability \cite{tufano_when_2017,palomba_diffuseness_2018}. Recent works have explored \llms to detect and refactor smells \cite{kaniewski_vulnerability_2024, hajipour_codelmsec_2024, siddiq_sallm_2024, yang_dlap_2024, tufano_when_2017, lucas2024evaluatinglargelanguagemodels, 10220911, li_multi-label_2022, di_wu_ismell_2024}, showing promising results when validated against ground truth. However, traditional metrics such as BLEU \cite{metric_bleu}, CodeBLEU \cite{ren_codebleu_2020}, ROUGE \cite{lin_rouge_2004}, and METEOR \cite{mwtric_meteor} fail to capture \llms' \textit{tendency} to introduce code smells at inference time. This tendency information is especially relevant for developers using commercial \llms (\eg ChatGPT, Copilot, and Claude\footnote{https://chatgpt.com, https://copilot.microsoft.com, https://claude.ai}), which may generate smelly snippets: \texttt{invalid-names}, \texttt{too-many-arguments}, or \texttt{unnecessary-lambda}. Although \llms are largely investigated in SE life-cycle \cite{10.1145/3485275}, little is known about the likelihood \llms generate smelly code.}

Despite the apparent success of using \llms for automating software tasks, developers still face two key challenges: 1) they need more information to assess which \llm is more reliable, and 2) they cannot evaluate model performance beyond accuracy, as traditional metrics often overestimate code quality. This leads to important questions: \textit{What is the propensity of an \llm to generate code smells?} and \textit{Which types of code smells are most propense to be generated?}

To address these concerns, we propose a new benchmark, \benchmark, designed to estimate the propensity of an \llm at generating smells. \benchmark draws inspiration from \textit{Syntax Decomposition} \cite{asttrust_2024, velasco_syntactic_capabilities} and introduces a model-agnostic evaluation metric called the \textit{Propensity Smelly Score (\metric)}. \metric analyzes the logits from the final layer of an \llm, providing insights into the model's propensity of generating code smells. Additionally, our benchmark includes a new dataset, \dataset, comprising $142k$ unique method-level code smells of $13$ different types, mined from GitHub.




To demonstrate the utility of our benchmark, we designed an exploratory case study using \benchmark in two \llms (\ie CodeLlama and Mistral). We assessed the distribution of \metric values for each type of smell in both models. Our analysis revealed that both models are propense to generate the same types of smell, with a few exceptions. For example, \textit{consider-merging-isinstance} (R1701), \textit{chained-comparison} (R1716), and \textit{broad-exception-caught} (W0718) were among the top, whereas \textit{disallowed-name} (C0104),  \textit{too-many-arguments} (R0913), and \textit{non-ascii-name} (C2401) were bellow the \metric propensity threshold. We hope that our findings will shed light on the propensity of current \llms to introduce code smells, enabling a more systematic and rigorous evaluation of code quality beyond canonical accuracy.


To summarize, our key contributions are as follows: 1) a new metric to evaluate the propensity of \llms to produce smells during code generation (\metric), 2) a new dataset \dataset of Python methods with \textit{Pylint} smells, 3) an exploratory case study to highlight the propensity of two current popular open-source \llms to produce smells, and 4) and notebooks and code packages with instructions to replicate the experiments and use our benchmark \cite{repository}. 

%% file: sections/2_benchmark.tex
\section{Benchmark}
\label{sec:benchmark}
In this section, we present our benchmark, \benchmark, which consists of three key components: (1) \metric, a metric designed to estimate the propensity of an \llm to introduce code smells, (2) our evaluation dataset, \dataset, comprising $142k$ curated instances of method-level code smells mined from GitHub, and (3) a \textit{protocol}, which outlines the methodology for using our benchmark.

\begin{figure*}[ht]
		\centering
  \vspace{-1em}
  \includegraphics[width=1\textwidth]{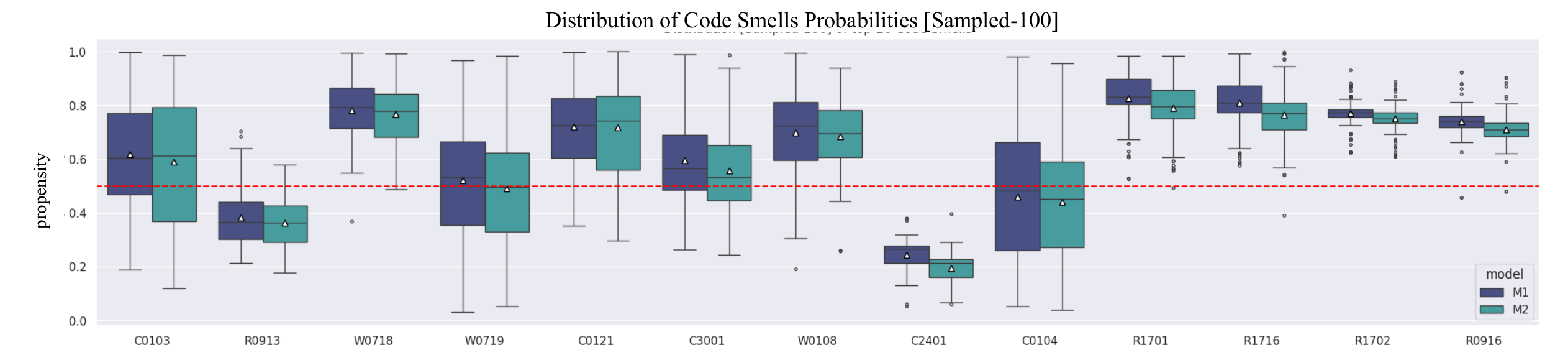}
		\caption{Bootstrapped propensity of smells (100 instances per sample). The red line at $0.5$ indicates the error threshold.}
    \label{fig:boxplots}
    \vspace{-1em}
\end{figure*}

\subsection{Propensity Smelly Score (\metric)}
\label{sec:benchmark_metric}


\metric is an evaluation metric that works by extracting the non-normalized log-probabilities (logits) $Z$ for each token prediction from the last hidden layer of a decoder-based transformer (\eg GPT). To estimate the probabilities of expected tokens in a sequence $w$, we apply the softmax function ($\sigma$) to each logit $z_i$. For a token $w_i = t$, where $t$ belongs to the vocabulary $V$, we compute the probability $P(w_i = t | w_{<i}) \approx \sigma(z_i)_t = {e^{z_{i,t}}}/{\sum_{j=1}^{|V|}{e^{z_{i,j}}}}$. In this equation, $z_{i,t}$ represents the logit for the expected token $t$ at position $i$, and the denominator normalizes the probabilities by summing the exponentiated logits for all vocabulary tokens. Since decoder-based models are auto-regressive, the preceding context influences the computation of $\sigma(z_i)_t$.


\textbf{Meaningful Structures $\mathbb{M}$.} Using an alignment function ($\delta$), tokens $w_i \in w$ are grouped into a meaningful structure $\mu \in \mathbb{M}$ (\equref{eq:alignment_function}). Then, an aggregation function ($\theta$) (\equref{eq:score_computation}) computes a central tendency statistic (\eg median, mean, mode) of their probabilities, resulting in an overall probability estimate for predicting $\mu$ (\ie propensity score). {Once $P(w_i | w_{<i})$ is estimated, the likelihood of each expected token is used to calculate the value of $\mu$. Each token position probability is treated as independent because NTPs (Next Token Predictions) are generated auto-regressively.}\equref{eq:score_computation} outlines how to compute the \metric for a meaningful concept $\mu$. Indexes pointing out smelly code are defined as $0 \leq i \leq k \leq j \leq |w|$.

\vspace{-0.2em}

\begin{equation}
\label{eq:alignment_function}
    \delta_\mu(w): w \to (i,j), \mu \in \mathbb{M}
\end{equation}

\begin{equation}
\label{eq:score_computation}
    \theta_\mu(w, i,j) =  \mathbb{E}_{k=i}^j[P(w_k | w_{0...k-1})]
\end{equation}

\begin{figure}[ht]
		\centering
  \includegraphics[width=0.5\textwidth]{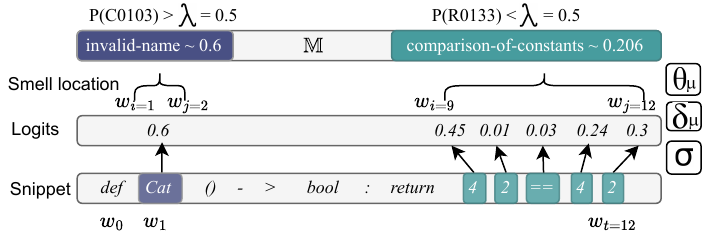}
		\caption{Example of \metric computation for invalid-name (C0103) and comparison-of-constants (R0133), with the former surpassing the propensity threshold $\lambda = 0.5$.}
    \label{fig:score_example}
\vspace{-1em}
\end{figure}

\textbf{Code Smells.} The definition of the set of meaningful structures $\mathbb{M}$ used in function $\delta$ (\equref{eq:alignment_function}) depends on the problem context. For example, token probabilities can be aggregated using syntax-based decomposition, based on elements defined by the grammar of a programming language (\eg identifiers, conditionals, statements). In this paper, we propose defining the set of concepts $\mathbb{M}$ using types of smells at method-level. \tabref{tab:code_smells} illustrates the full taxonomy of code smells in $\mathbb{M}$ included in \dataset.

\textbf{Global Estimates.} \label{sec:benchmark_metric_global} Finally, we compute the average of \equref{eq:score_computation} for a given code smell type ($\mu$) across all code snippets $s$ in the dataset (\dataset) to estimate the overall \metric of code smells generated by an \llm. We define a \textit{propensity threshold} $\lambda = 0.5$ to determine whether the \llm is \textit{propense} to generate the code smell, with $\metric \geq \lambda = 0.5$ indicating a higher propensity. We use this estimate to answer \ref{rq:code_smell_propensity}. {The propensity threshold, based on the work of Karpaty \etal~\cite{karpathy_visualizing_2015}, is set at $0.5$ to indicate that code smells are more likely to be produced than flipping a coin.} In addition, \figref{fig:score_example} depicts an example of computing \metric for a Python snippet containing the code smells: \textit{invalid-name} with a \metric of $0.6$ above the propensity threshold and \textit{comparison-of-constants} with a \metric of $0.206$. 

\subsection{\dataset}

\input{tables/smells}

To mitigate the risk of data leakage during the smells evaluation, we created a new testbed: \dataset using Galeras \cite{rodriguez-cardenas_benchmarking_2023}. We mined popular open-source Python repositories from GitHub, extracting a total of $232,715$ methods. The selection criteria included repositories published between January 2022 and December 2024, with a minimum of $3,500$ stars, ensuring that only well-maintained and highly rated repositories were considered. We then applied \textit{Pylint} \cite{pylint_messages} analysis to these methods, detecting $79,574$ with at least one code smell.


\textbf{Dataset Curation.} From this filtered set of methods, we identified $156,151$ instances of smells across $30$ distinct types. Using random sampling, two authors manually validated instances from each code smell type to confirm true positives (TP) and false positives (FP) with an $80\%$ confidence level and $15\%$ margin error. {Discrepancies in the validation results were resolved through discussions among the authors}. We found that five code smells—\textit{inconsistent-return-statements}, \textit{too-many-branches}, \textit{too-many-return-statements}, \textit{too-many-statements}, and \textit{unbalanced-tuple-unpacking}—had near to zero precision. For \textit{unbalanced-tuple-unpacking}, the number of FPs exceeded the number of TPs, while for the other cases, \textit{Pylint} failed to identify the exact location on the other code smell types of the smell within the method due to parsing errors. Additionally, $12$ smells were excluded because they had fewer than $100$ instances, making them not representative (we observed that all \metric distributions tend to resemble a Gaussian distribution with at least $100$ instances). Finally, we ensured that all code smell instances in \dataset originated from unique Python methods, avoiding any repetition of source code among code smell types. The resulting dataset consists of $142,817$ code smell instances from $13$ confirmed types, categorized into Refactoring, Warnings, and Conventions, as shown in \tabref{tab:code_smells}.

\textbf{Code Smell Location.} In our case study, we used the information provided by \textit{Pylint} (\ie starting and ending row and column) to locate the exact portion of each code snippet containing a code smell, which allowed us to implement the alignment function $\delta$ as introduced in \equref{eq:alignment_function}.

\subsection{Protocol}
\label{sec:benchmark_protocol}
\metric evaluates the propensity of \llms to generate specific types of code structures by analyzing next-token predictions. Our benchmark computes the probability for each expected token in the sequences from \dataset. The steps are as follows: (1) encode each sequence using the \llm's tokenizer, (2) compute the logits for each token using the \llm's forward method, (3) apply the softmax function ($\sigma$) to calculate the probability of each expected value, (4) group tokens corresponding to a given code smell using the alignment function ($\delta$), and (5) compute the \metric for each code smell using a central tendency statistic (\equref{eq:score_computation}). The resulting \metric estimates the propensity of each smell to be generated by an \llm.

%% file: tables/smells.tex
\begin{table}[]

\centering
\caption{Method-level code smells instances included in \dataset.}
\label{tab:code_smells}
\vspace{-0.5em}
\scalebox{1}{

\setlength{\tabcolsep}{4pt} 

\begin{tabular}{llll}
\textbf{Type}                        & \textbf{ID} & \textbf{Code Smell}    $\mu \in \mathbb{M}$                 & \textbf{Count} \\ \hline
\multirow{5}{*}{\textit{Convention}} & C0103       & \textit{invalid-name}                  & 125815         \\
                                     & C0121       & \textit{singleton-comparison}          & 1089           \\
                                     & C3001       & \textit{unnecessary-lambda-assignment} & 666            \\
                                     & C2401       & \textit{non-ascii-name}                & 583            \\
                                     & C0104       & \textit{disallowed-name}               & 174            \\
                                     &             & \textit{}                              &                \\
\multirow{5}{*}{\textit{Refactor}}   & R0913       & \textit{too-many-arguments}            & 4738           \\
                                     & R1702       & \textit{too-many-nested-blocks}        & 1273           \\
                                     & R0916       & \textit{too-many-boolean-expressions}  & 289            \\
                                     & R1701       & \textit{consider-merging-isinstance}   & 132            \\
                                     & R1716       & \textit{chained-comparison}            & 128            \\
                                     &             & \textit{}                              &                \\
\multirow{3}{*}{Warning}             & W0718       & \textit{broad-exception-caught}        & 4384           \\
                                     & W0719       & \textit{broad-exception-raised}        & 3150           \\
                                     & W0108       & \textit{unnecessary-lambda}            & 396            \\ \hline
\end{tabular}
} 
\caption*{\small{*Refer to \textit{Pylint} \cite{pylint_messages} for detailed smells' description.}}
\vspace{-1.5em}
\end{table}

%% file: sections/3_case_study.tex
\section{Case Study}
\label{sec:case_study}

To demonstrate the practical application of our benchmark, we conducted a case study evaluating the propensity of two \llms to generate the code smells in \dataset. We formulated the following main research question:

\begin{enumerate}[label=\textbf{RQ$_{\arabic*}$}, ref=\textbf{RQ$_{\arabic*}$}, wide, labelindent=5pt]\setlength{\itemsep}{0.2em}
      \item \label{rq:code_smell_propensity} {\textbf{[Propensity of Code Smells]} What types of code smells are more propense to be generated by M1 and M2?}
\end{enumerate}

\textbf{Selected \llms.} Although \metric is model-agnostic, we selected two popular decoder-based transformers as they are well-suited for generative tasks. The first model, CodeLlama-7b-Instruct-hf (M1) \cite{roziere_code_2024}, has a vocabulary size of $32,016$ tokens. The second model, Mistral-7B-v0.3 (M2) \cite{jiang_mistral_2023}, has a vocabulary size of $37,768$ tokens. Both models have 7 billion parameters, 32 hidden layers, and 32 attention heads. The models were loaded on an Ubuntu 20.04 system with an AMD EPYC 7532 32-Core CPU, an NVIDIA A100 GPU with 40GB VRAM, and 1TB of RAM.

\textbf{Evaluation Methodology}. To address \ref{rq:code_smell_propensity}, we computed \metric global estimates (refer to \secref{sec:benchmark_metric_global}) for both models (\ie M1 and M2) using the collected snippets for all identified code smells. Due to memory constraints, we limited the evaluation to datapoints in \dataset with a maximum size of $400$ tokens. We further sampled $100$ datapoints from each of the $13$ code smells with at least $100$ instances in the dataset to ensure fair statistical treatment. Also, note that all the selected datapoints came from distinct Python methods. We then followed the protocol steps (refer to \secref{sec:benchmark_protocol}) to compute the global estimates of \metric for each code smell.

\subsection{Results \& Discussion}
\label{sec:case_study_results_discussion}

\input{tables/short_results}


\figref{fig:boxplots} presents the probability scores computed for each code smell using M1 and M2. Remarkably, $10$ out of $13$ code smells have a probability higher than the propensity threshold of $0.5$, indicating that both models are propense to generating these smells. As shown in \tabref{tab:short_results}, the top five smells with the highest \metric in both models are \textit{consider-merging-isinstance} (R1701), \textit{chained-comparison} (R1716), \textit{broad-exception-caught} (W0718), \textit{too-many-nested-blocks} (R1702) and \textit{too-many-boolean-expressions} (R0916). Conversely, \textit{disallowed-name} (C0104), \textit{too-many-arguments} (R0913) and \textit{non-ascii-name} (C2401) are bellow the propensity threshold for both models. 

\begin{figure}[ht]
		\centering
  \vspace{-1.0em}
  \includegraphics[width=0.5\textwidth]{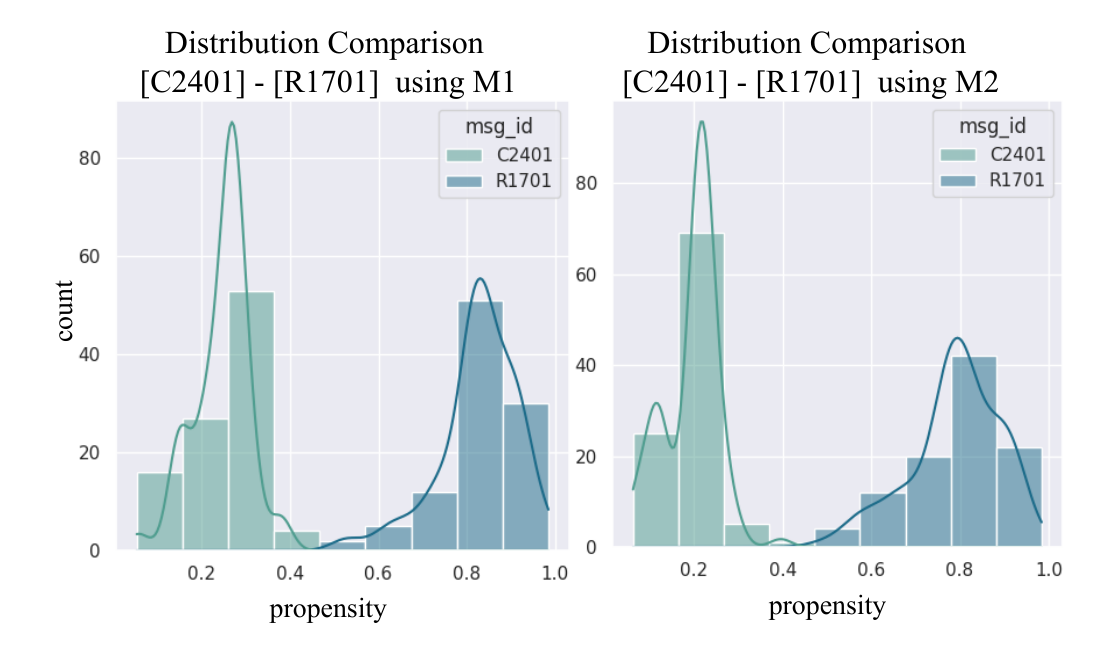}
		\caption{Comparison of edge propensity distributions between (\textit{consider-merging-isinstance} - R1701) and (\textit{disallowed-name} - C2401) for both M1 and M2.}
    \label{fig:dist_comparison}
    \vspace{-1.0em}
\end{figure}

Upon examining the computed distributions of \metric for both the highest and lowest scoring code smells in each model, as shown in \figref{fig:dist_comparison}, we observe a significant difference. The \metric score for \textit{consider-merging-isinstance} (R1701) is nearly double that of \textit{disallowed-name} (C2401). We attribute this disparity to the fact that tokens associated with disallowed names have very low probabilities, as they are highly specific and closely tied to the context of the source code.

\begin{boxK}
    \textit{\ref{rq:code_smell_propensity}} \textbf{[Propensity of Code Smells]}: Both CodeLlama-7b-Instruct-hf \cite{roziere_code_2024} and Mistral-7B-v0.3 \cite{jiang_mistral_2023} are propense to generate code-smells Convention, Refactor, and Warnings. With few exceptions: \textit{disallowed-name} (C0104),  \textit{too-many-arguments} (R0913), and \textit{non-ascii-name} (C2401).
\end{boxK}

We believe that both \llms are propense to generate detected code smells, as they are trained on publicly available code, which often contains quality issues~\cite{roziere_code_2024}. Moreover, no evidence suggests that the training data of the selected \llms has been curated to eliminate smells. This assumption is reinforced by our dataset creation process since we extracted smells from mined code snippets in public repositories. Nevertheless, a separate study is required to confirm this hypothesis as our goal is to demonstrate the utility of \benchmark. 

The results indicate that higher \metric values are correlated with a higher tendency to produce code smells, potentially reflecting models' bias toward specific types of smells. However, since \metric is computed from conditioned generation (\ie from code snippets containing a code smell), another study is needed to determine whether a correlation exists between \metric scores and the overall quality of the code generated by \llms. While our observations highlight the need for larger-scale experiments and further validation, a detailed investigation into the underlying causes of code smells in generated code is beyond the scope of this paper. 

%% file: tables/short_results.tex
\begin{table}[]

\centering
\caption{Top-5 highest and Top-3 lowest (gray bkg.) smells ranked by \metric (in [avg$\pm$std]).}
\label{tab:short_results}
\vspace{-0.5em}
\scalebox{1}{

\setlength{\tabcolsep}{4pt} 

\begin{tabular}{llclclc}
\multicolumn{1}{c}{\textbf{Code Smell}} &  & \textbf{M1 \metric} &  & \textbf{M2 \metric} &  & \multicolumn{1}{l}{\textbf{ME - $95\%$} } \\ \hline
R1701 &  & 0.80 $\pm$ 0.08  &  & 0.80 $\pm$ 0.10   &  & $5\%$  \\
R1716 &  & 0.80 $\pm$ 0.10   &  & 0.77 $\pm$ 0.10  &  & $5\%$  \\
W0718 &  & 0.80 $\pm$ 0.12  &  & 0.77 $\pm$ 0.10  &  & $10\%$ \\
R1702 &  & 0.77 $\pm$ 0.05 &  & 0.75 $\pm$ 0.06 &  & $10\%$ \\
R0916 &  & 0.73 $\pm$ 0.06 &  & 0.71 $\pm$ 0.06 &  & $8\%$  \\
\rowcolor[HTML]{EFEFEF} 
C0104 &  & 0.46 $\pm$ 0.25 &  & 0.44 $\pm$ 0.23 &  & $7\%$  \\
\rowcolor[HTML]{EFEFEF} 
R0913 &  & 0.40 $\pm$ 0.13  &  & 0.36 $\pm$ 0.10  &  & $10\%$ \\
\rowcolor[HTML]{EFEFEF} 
C2401 &  & 0.24 $\pm$ 0.07 &  & 0.20 $\pm$ 0.06  &  & $9\%$  \\ \hline
\end{tabular}

} 
\caption*{\small{*ME - Margin Error for sample size of $100$.}}
\vspace{-1.5em}
\end{table}

%% file: sections/4_related.tex
\section{Related Work}\label{sec:background}

Considerable research has been devoted to collecting code smell data, detecting, and repairing. Code smell often indicates deeper issues in the codebase, affecting code quality and performance~\cite{mahalakshmi_code_2023, siksna_machine_2023,mohsin_can_2024,hajipour_codelmsec_2024,siddiq_sallm_2024,kaniewski_vulnerability_2024,palomba_diffuseness_2018,tufano_when_2017}. Our related work is focused on research on code smell datasets and papers that have reported benchmarks on generating code smells.



Nasrabadi \etal \cite{zakeri-nasrabadi_systematic_2023}  introduced an SLR with the most updated code smells datasets. Most datasets are training or testing code smell detection on \llms ~\cite{mohsin_can_2024,hajipour_codelmsec_2024, nguyen_thanh_ml-codesmell_2022}. Datasets can be created manually \cite{Madeyski2020MLCQIC,Hozano2017EvaluatingTA,nandani_dacosmanually_2023} or automatically \cite{lenarduzzi_technical_2019,8377639}
via refactorings, however, the validations are time-consuming so most automatically generated datasets are not validated\cite{zakeri-nasrabadi_systematic_2023}. Our dataset \dataset is automatically mined and manually curated to confirm the code smell type and location.


Existing benchmarks focus on detecting code smells using different techniques (\eg SVM~\cite{nguyen_thanh_ml-codesmell_2022}, few-shot learning~\cite{hajipour_codelmsec_2024}, chain-of-thoughts~\cite{kaniewski_vulnerability_2024}). CodeLMSec~\cite{hajipour_codelmsec_2024} evaluates code generation models' vulnerability to generating insecure prompts, while LCG~\cite{lin_when_2024}, iSmell~\cite{di_wu_ismell_2024}, and PromptSmell~\cite{liu_prompt_2024} examine \llm-based approaches. Unlike these, our benchmark uses logits to assess an \llm's propensity for generating smelly code, rather than merely classifying or detecting it. By aligning meaningful tokens to highlight smelly segments, our approach offers a novel, interpretable metric~\cite{rodriguez-cardenas_benchmarking_2023},~\cite{nader_palacio_toward_2024}, suitable for black-box code generation models.

%% file: sections/5_future_plan.tex
\section{Future Plans}
\label{sec:future_plan}


Based on the results of our case study, we demonstrated the computation of \metric for $13$ method-level code smells. {As a next step, we plan to conduct systematic experiments to evaluate the robustness of \metric by incorporating a wider variety of smells and a broader range of \llms}. This effort will involve mining more instances of underrepresented code smells in \dataset. Specifically, we will analyze sampling error when computing \metric for each code smell to mitigate the risk of sampling bias and ensure the representativeness of all smell types in \dataset. Furthermore, since our analysis has focused solely on method-level granularity, we plan to expand \dataset to include higher-level smells, such as \textit{god-class} and \textit{feature-envy}. In future versions of \dataset, we will use other static analysis tools alongside \textit{Pylint} to reduce the dependency on a single tool for identifying smell locations, which is crucial to implement the alignment function (\equref{eq:alignment_function}) in \benchmark.

{Our proposed benchmark identifies the types of code smells that \llms are propense to generate. However, we acknowledge the need for empirical validation to demonstrate that \textit{PSC} scores align with real-world developer expectations and impact. Future research should address how \textit{PSC} supports practitioners in interpreting a model's behavior and how it can be effectively used in practice. For instance, \textit{PSC} could provide actionable insights by highlighting the types of smells a model is likely to introduce, enabling developers to assess model outputs more critically and prioritize mitigation strategies. Additionally, future work should uncover the reasons behind \llms' propensity to introduce specific types of smells, potentially by identifying the most relevant input features that influence the generation of these smells. This research direction is crucial for improving the trustworthiness~\cite{asttrust_2024} of \llms and developing defense techniques to mitigate such behaviors. We believe interpretability techniques such as LIME~\cite{lime_ribeiro} or SHAP~\cite{shap_lundberg} will be instrumental in achieving these goals. 
